\documentclass[12pt]{JHEP3}
\usepackage{epsfig}
\preprint{ {\tt hep-th/0305217}} 
\newcommand{\be}[1]{ \begin{equation}\label{#1} }
\newcommand{\ee}{\end{equation}}
\newcommand{\bea}[1]{\begin{eqnarray}\label{#1} }
\newcommand{\eea}{\end{eqnarray}}
\newcommand{\eq}[1]{(\ref{#1})}

\title{Gravitational F-terms through
anomaly equations and deformed chiral rings}
\author{Luis F. Alday$^{a,b}$, 
Michele Cirafici$^b$, 
Justin R. David$^a$ , Edi Gava$^{a, b}$, K.S. Narain$^{a}$ \\
$^a$High Energy Section, \\
 The Abdus Salam International Centre for Theoretical Physics, 
\\Strada Costiera, 11-34014 Trieste, Italy.\\
$^b$Instituto Nazionale di Fisica Nucleare, sez. di Trieste, \\
and SISSA, Italy. \\
\email{alday@he.sissa.it, cirafici@sissa.it, \\
justin, gava, narain@ictp.trieste.it}
}

\abstract{
We study effective gravitational F-terms,
obtained by integrating an $U(N)$ adjoint chiral superfield $\Phi$
coupled to the ${\cal N}=1$ gauge chiral superfield $W_\alpha$
and supergravity, to arbitrary
orders in the gravitational background. The latter
includes in addition to the ${\cal N}=1$ Weyl
superfield $G_{\alpha\beta\gamma}$, the self-dual
graviphoton field strength $F_{\alpha\beta}$ of the
parent, broken ${\cal N}=2$ theory. We first study the chiral
ring relations resulting from the above non-standard  gravitational
background and find agreement, for gauge invariant operators,
with those obtained from the dual closed string side via
Bianchi identities for ${\cal N}=2$ supergravity coupled to vector
multiplets.
We then derive generalized anomaly equations for connected correlators
on the gauge theory side, which allow us to solve for the basic
one-point function $\langle {\rm Tr}  W^2/(z-\Phi)\rangle$ to all orders in 
$F^2$.
By generalizing the matrix model loop equation to
the generating functional of connected correlators of resolvents,
we prove that the gauge theory result coincides with the genus 
expansion of the associated matrix model, after identifying the
expansion parameters on the two sides.}

\begin{document}
\baselineskip 4ex

\section{Introduction}

The issue of effective gravitational F-terms in
${\cal N}=1$ gauge theories coupled to ${\cal N}=1$ supergravity, generated
by integrating out a massive $U(N)$ adjoint superfield $\Phi$,
has received much attention recently
\cite{Dijkgraaf:2002dh,Klemm:2002pa,Dijkgraaf:2002yn,
Ooguri:2003qp,Ooguri:2003tt,David:2003ke}, 
after the conjecture of \cite{Dijkgraaf:2002dh}, 
which has identified these terms
to higher genus contributions to the free energy of a related
hermitian matrix model, with potential given by the superpotential
$W(\Phi)$ of the adjoint superfield. The conjecture was proved
in \cite{Ooguri:2003qp,Ooguri:2003tt} 
using the diagrammatic techniques developed
in \cite{Dijkgraaf:2002xd} 
to prove the relation between the effective
superpotential of the gauge theory and the planar, genus zero,
contribution to the matrix model free energy.
One of the ingredients of the proof in 
\cite{Ooguri:2003qp,Ooguri:2003tt} was
the modification of the chiral ring of the gauge theory in
the presence of a supergravitational background. The modification
has two sources. The first can be understood within the framework
of ordinary supergravity tensor calculus and related Bianchi
identities in the presence of the ${\cal N}=1$ Weyl superfield
$G_{\alpha\beta\gamma}$, and it reads $\{W_\alpha, W_\beta \}=
2G_{\alpha\beta\gamma}W^\gamma$, where $W_\alpha$ is the
gauge chiral superfield. It turns out however that, in order
to capture genus $g\geq 2$ contributions,
in the language of the dual closed
string theory side \cite{Vafa:2000wi}, 
one needs to introduce a more
drastic modification in the chiral ring relation, to account
for a non-trivial vacuum expectation value 
of the (self-dual) graviphoton field
strength $F_{\alpha\beta}$ of the parent ${\cal N}=2$ string theory.
In this case the relation reads
$\{W_\alpha, W_\beta \}= F_{\alpha\beta}+
2G_{\alpha\beta\gamma}W^\gamma$. In particular, one is
modifying the Grassmann nature of the fermionic superfield
$W_\alpha$.  In \cite{David:2003ke} 
we considered the first kind of modification,
involving only the Weyl superfield $G_{\alpha\beta\gamma}$,
exploited its consequences and used generalized anomaly
equations\cite{Cachazo:2002ry}  to prove that, indeed, the order $G^2$
effective superpotential is given, as a function of
$S= - {\rm Tr}  W^2/32\pi^2$, by the genus one contribution to the
free energy of the related matrix model. A crucial fact in
that approach was the lack of factorization of chiral
correlators in the presence of a non-trivial supergravity
background, in particular the non-triviality of connected
two-point correlators.

The purpose of the present work is to complete and extend the above
analysis to arbitrary orders in $F^2$, i.e. to
arbitrary genera on the matrix model side, or on the
dual closed string side. This amounts, among other things,
to determine the modifications in the chiral ring
relations due to the presence of  the $F_{\alpha\beta}$ 
background. It turns out that in this case one has to face
ordering ambiguities in  manipulating $W$'s in the generalized
Konishi anomaly equations, somewhat similar to the base point
dependence in path ordered exponentials found in 
\cite{Ooguri:2003qp,Ooguri:2003tt}.
We fix the ambiguities by requiring that
traces of (graded) commutators be trivial in the chiral ring.
This requirement will lead us to get identities involving gauge
invariant and gravitational operators, which
will be very important
when analyzing the anomaly equations. While we do not have a
proof of this assumption at the gauge theory level, quite remarkably
we will find that these relations are exactly the same as those
we obtain on the dual closed string theory side by using the Bianchi
identities of the (broken) ${\cal N}=2$ supergravity theory coupled to
vector multiplets. We will then first, following the strategy of
\cite{David:2003ke},
reconsider the $g=1$ case in the presence of both
$F_{\alpha\beta}$ and $G_{\alpha\beta\gamma}$ and find that the
order $F^2$ and $G^2$ superpotential terms have the
expected structure and agree with the $g=1$ matrix model
result.

In order to proceed with arbitrary genus analysis, for
which a non-trivial $F_{\alpha\beta}$ is essential,
we will need to further generalize the strategy of 
\cite{David:2003ke}
to derive anomaly equations for the generating functional
of (connected) correlators, by coupling the relevant
chiral gauge invariant operators to external sources.
From these equations we will extract the
correlators which are required to determine the
one-point functions of $R(z)$,
$T(z)$
and  $w_\alpha(z)$ (in the notation of \cite{Cachazo:2002ry}
reviewed in section 2)
to all orders in $F^2$. From these
one can determine the effective superpotential as
explained in \cite{Cachazo:2002ry}. 
The connection with the matrix
model will be proved by generalizing the loop equation
of the latter to a full generating functional of connected
correlators of resolvents, by coupling the matrix model
resolvent $\Omega_m (z)$ to an external source. 
We will prove that, in fact, the gauge theory $\langle R(z)\rangle$
coincides with the matrix model $\langle \Omega_m(z)\rangle$,
to all orders, if we identify $F^2$ with the matrix model
genus-expansion parameter $1/{\hat N}^2$.
More precisely, there is a full class of gauge theory correlators,
$\langle {\rm Tr} W^2 {\Phi}^k \rangle$, whose expansion in powers of $F^2$
coincides with the $1/{\hat N}^2$ expansion of $\langle 
{\rm Tr}M^k \rangle$, where $M$ is a ${\hat N}
\times {\hat N}$ hermitian random matrix, whose expectation 
value is computed with the measure 
$\exp(-\frac{\hat N}{g_m}{\rm Tr}W(M))$. 
We find this fact quite remarkable.

It is also worth mentioning that
lower genus contributions to a given genus term can be gotten
rid of by an operator redefinition, $R\rightarrow R+\frac{1}{6}
G^2 T$, thereby generalizing the shift $S\rightarrow S+\frac{1}{6}
G^2 N$ needed to remove the genus zero contribution to
the order $G^2$ term.

It is interesting to note that, by following
\cite{Cachazo:2002ry} 
and introducing an auxiliary
Grassman coordinate $\psi_\alpha$, which can be thought
of as a second supercoordinate of the broken ${\cal N}=2$ parent theory,
we can rewrite the anomaly equations in a shift-invariant way,
if, in addition to assembling $R$, $w_\alpha$ and $T$
in ${\cal R}(\psi)$, as in \cite{Cachazo:2002ry},
we assemble also $F$ and $G$ as ${\cal H}=F_{\alpha\beta}-
\frac{1}{2}\psi^\gamma G_{\alpha\beta\gamma}$. This suggests
that the  effective
superpotential can be formally written in a manifest shift-invariant
way, $\int d^2\psi {\cal H}^{2g}{\cal F}_g(S+\psi^\alpha w_\alpha
-\frac{1}{2}\psi^2 N)$. However, it will turn out that,
due to the chiral ring relations, to be derived in section 2,
the $g\geq 2$ terms are all trivial from the ${\cal N}=1$ point of
view. We should stress, nevertheless once again, that the all-orders
identification of the gauge theory $\langle R(z)\rangle$ 
with the matrix model $\langle \Omega_m(z)\rangle$, implying
the exact identity of an infinite family of correlators on the two
sides, survives the chiral ring relations, since on the gauge theory
side it involves powers of $F^2$ only. 

The paper is organized as follows: in section 2 we discuss the
chiral ring relations in the gauge theory with background
$F$ and $G$ and those coming from the dual closed string side.
In section 3 we consider the genus one contribution, extending
the discussion of \cite{David:2003ke} to the case where both $F$ and
$G$ are non-trivial and proving the expected structure
of order $G^2$ and $F^2$ terms. In section 4 we derive
anomaly equations for the generating functional of connected
correlators, find the relevant connected two-point functions
and finally the all orders one point function of $T$, from
which the superpotential can be integrated. We also verify the
shift symmetry of our expressions and prove agreement
with the matrix model result. In appendix A we prove the
consistency of the anomaly equations for the generating functional.

\section{The chiral ring}

The chiral ring in ${\cal N} =1$ gauge theories consists of all operators
which are annihilated by the covariant derivative $\bar{D}_{\dot{\alpha}}$, 
which is conjugate 
to the supercharge $\bar{Q}_{\dot{\alpha}}$, modulo
$\bar{D}_{\dot{\alpha}}$ exact terms, where the exact terms should be gauge
invariant and local operators.
All relations in the chiral ring are therefore defined  modulo
$\bar{D}_{\dot{\alpha}}$ exact terms. 
The chiral ring and the various relations in it 
play an important role in deriving the effective
$F$-terms in ${\cal N}=1$ gauge theory obtained by integrating out the
adjoint matter \cite{Cachazo:2002ry}. 
This is because 
the contributions to the effective action 
of all $\bar{D}_{\dot{\alpha}}$ exact terms 
can be written as an integral involving both holomorphic and
anti-holomorphic integrations in superspace, $\int d^2x d^2\theta
d^2\bar{\theta} S(\theta, \bar{\theta})$. Thus they do not contribute
to the F-terms, which involve only either a holomorphic or
anti-holomorphic integrations in the superspace co-ordinates.

For the ${\cal N}=1$ gauge theory with a single adjoint field 
on $R^4$ the chiral
ring relations are given by \cite{Cachazo:2002ry} \footnote{All
relations in the chiral ring are modulo $\bar{D}$, for 
the rest of paper this will not be explicitly mentioned, but
understood wherever necessary.}
\be{oring}
[W_\alpha, \Phi] =0 \; \hbox{mod}\; \bar{D}, \;\;\;\; \{W_\alpha,
W_\beta\} = 0 \; \hbox{mod}\; \bar{D}, 
\ee
where $W_\alpha$ is the ${\cal N}=1$ gauge chiral superfield
and $\Phi$ is the chiral
matter superfield in the adjoint representation of the gauge group.
In this paper we are interested in studying the F-terms in ${\cal N}=1$
$U(N)$ gauge theory 
obtained by integrating out a single adjoint scalar in presence of
gravity as well as the self-dual graviphoton field strength
$F_{\alpha\beta}$, using the generalized anomaly equations approach. 
The chiral ring relations in the presence of these backgrounds are given
by \cite{Ooguri:2003qp,Ooguri:2003tt}
\be{ring}
[W_\alpha, \Phi] =0, \;  \;\;\;\; \{W_\alpha,
W_\beta\} =F_{\alpha\beta} + 2G_{\alpha\beta\gamma}W^\gamma. 
\ee
In the presence of just a curved background the ring relations can be
derived by using the Bianchi identities of ${\cal N}=1$ gravity, but for
obtaining the ring relation in the presence of the graviphoton one has to
resort to string theory. It was pointed out in
\cite{Ooguri:2003qp} that
the modification of ring relations in the presence of $F_{\alpha\beta}$
requires a non-traditional interpretation. The classical Grassmanian
nature of the  ${\cal N}=1$ gauge multiplet 
no longer holds, the gauge multiplet in fact satisfies a Clifford
algebra. 
All the gauge invariant operators constructed out of the basic 
fields of the ${\cal N}=1$ gauge multiplet and the chiral multiplet
can be arranged into the following operators
\bea{defoper}
{\cal R}(z)_{ij} = -\frac{1}{32\pi^2} \left( \frac{W^2}{z-\Phi}
\right)_{ij},   &\;& R(z) = {\rm Tr } {\cal R}(z), \\ \nonumber
\rho_\alpha (z)_{ij}  = \frac{1}{4\pi} \left( \frac{W_\alpha}{z-\Phi}
\right)_{ij}, &\;& w_\alpha(z) = {\rm Tr} \rho_\alpha (z), \\
\nonumber
{\cal T}(z)_{ij} = \left( \frac{1}{z-\Phi} \right)_{ij}, &\;&
T(z) = {\rm Tr} {\cal T} (z). 
\eea
Here, for later convenience,  
we have defined separate symbols for the matrix elements operators
and the gauge invariant, trace operators. Placing more $W$'s in the
trace does not yield any more gauge invariant operators, as they can be
converted to one of the above operators by the ring relations in \eq{ring},
therefore the above set of operators is exhaustive in the chiral ring.
In the next sub-section we will derive various identities from the
relations \eq{ring}, with a motivated ansatz that the adjoint action with
$W_\alpha$ on gauge invariant operators vanishes. 
In the subsequent sub-section we will re derive these identities from
the closed string dual using the ${\cal N}=2$ Bianchi identities, thus
justifying the ansatz.

\subsection{Chiral ring identities from gauge theory}

From the definition of $W^2$ and the basic ring equations in \eq{ring}
we can derive the following identities using simple algebraic
manipulations,
\bea{comm}
[W_\alpha, W^2] &=& -2 F_{\alpha\beta} W^\beta, \\ \nonumber
\{W_\alpha, W^2\} &=& -\frac{2}{3} \left( G^2 W_\alpha +
G_{\alpha\beta\gamma} F^{\beta\gamma} \right),
\eea
adding the above equations we find
\be{rrel1}
W_\alpha W^2 = -F_{\alpha\beta} W^\beta - \frac{1}{3} G^2 W_\alpha
-\frac{1}{3} G_{\alpha\beta\gamma} F^{\beta\gamma}.
\ee
Considering the  equation \eq{ring} with a product of arbitrary number of
scalars $\Phi$ and taking trace on both sides of the equation, we obtain
\bea{ringsc}
{\rm{Tr}}
( \{W_\alpha, W_\beta\} \Phi \Phi \ldots ) &=& 
{\rm{Tr}}( (F_{\alpha\beta} +2
G_{\alpha\beta\gamma} W^\gamma ) \Phi \Phi \ldots ), \\ \nonumber
&=& {\rm{Tr}}( {\rm{Ad}}_{ W_\alpha} W_\beta \Phi\Phi \ldots ).
\eea
We see that on the left hand side of the equation we have the adjoint
action on the operator $W_\beta \Phi \Phi \ldots$. For ordinary
Grassman $W_\alpha$, the trace of the adjoint action 
is zero, but here, from
the algebra in \eq{ring}, it is not clear that 
this will still hold. However if
the trace of the adjoint action is not zero, then in any gauge
invariant operator, like the one considered in \eq{ringsc}, there will
be an ambiguity in the ordering of $W_\alpha$,  such that the cyclic
property of the trace will not be obeyed. This is the base point
ambiguity noted in \cite{Ooguri:2003qp}. To remove such ambiguities we
demand that the trace of the adjoint action is trivial in the chiral
ring. This leads us to the following equation
\be{ringsc1}
{\rm Tr}( ( F_{\alpha\beta} + 2G_{\alpha\beta\gamma} W^\gamma) \Phi
\ldots ) = 0 .  
\ee
We can write the above equation compactly in terms of the operators
in \eq{defoper}. After performing the following convenient
redefinitions, 
$F_{\alpha\beta} \rightarrow 32\pi^2 2\sqrt{2} F_{\alpha\beta}$ and
$G_{\alpha\beta\gamma} \rightarrow \sqrt{32\pi^2} \
G_{\alpha\beta\gamma}$, \eq{ringsc1} can be written as
\footnote{From now on we will use these scaled variables for the rest
of the paper.}
\be{ring1}
2 F_{\alpha\beta} T+  G_{\alpha\beta\gamma} w^\gamma =0. 
\ee
A similar equation can be obtained by considering  the first equation in
\eq{comm} with products of $\Phi$'s and  a trace on both sides,
we obtain
\be{commsc}
{\rm Tr} ( [W_\alpha, W^2 ] \Phi\ldots ) = - 2F_{\alpha\beta} {\rm
Tr}( W^\beta \Phi \ldots ). 
\ee
Again demanding that the adjoint action is trivial in the chiral ring
we get
\be{commsc1}
F_{\alpha\beta} {\rm Tr} ( W^\beta \Phi \ldots) = 0,
\ee
that written  in terms of the operators in \eq{defoper} becomes
\be{ring2}
F_{\alpha\beta} w^\beta = 0. 
\ee

Using Bianchi identities of ${\cal N}=1$ supergravity
it was shown in \cite{David:2003ke}, 
that the spin 2 combination of a product of two ${\cal N}=1$
Weyl multiplet was trivial in the ring. This combination is given by
\be{gg}
G_{\alpha\beta\sigma} G^\sigma_{\;\;\gamma\delta} +
G_{\alpha\gamma\sigma} G^\sigma_{\;\;\beta\delta} +
G_{\alpha\delta\sigma} G^\sigma_{\;\;\beta\gamma} =0. 
\ee
This ensures that the gravitational corrections to the F-terms
truncate at order $G^2$.
If a similar Bianchi identity were applied on the symmetric tensor
$F_{\alpha\beta}$ we would obtain the following  product
of the graviphoton and the ${\cal N}=1$ Weyl multiplet
\be{gfp}
G_{\alpha\beta\gamma}F^{\gamma}_{\;\;\sigma} +
G_{\alpha\sigma\gamma}F^\gamma_{\;\;\beta}, 
\ee
which should therefore  vanish in
the chiral ring. 
However, a consistent ${\cal N}=1$ gauge theory coupled to the spin $3/2$
multiplet containing the graviphoton along with its supersymmetric
partner, the gravitino, is at present not known. Such theories are
consistent from string theory point of view as these are theories on
D-branes with the graviphoton field strength turned on in the bulk.
The presence of the graviphoton field strength would perhaps modify
the Bianchi identity in \eq{gfp},
without an explicit construction of the gauge theory 
coupled to graviphoton it is not possible to
determine the modification of the Bianchi identity on the $F_{\alpha\beta}$.
The  product of $G$ and $F$ in \eq{gfp} 
contains both a spin $3/2$ and a spin $1/2$ part, but we make the minimal
ansatz  that the spin $3/2$ combination in the tensor product is
trivial in the chiral ring  as given below
\be{gf}
G_{\alpha\beta\gamma}F^{\gamma}_{\;\sigma} +
G_{\alpha\sigma\gamma}F^{\gamma}_{\;\beta} +
G_{\sigma\beta\gamma}F^{\gamma}_{\;\alpha} = 0 
\ee

From the basic identities in the chiral ring, 
\eq{ring1}, \eq{ring2}, \eq{gg} and \eq{gf}, we derive 
other identities which are used at several instances in  this
paper. The first identity is obtained by
multiplying the equation \eq{ring2} by $F_{\alpha '}^
{\;\;\alpha}$, which is gives
\be{rel1}
F_{\alpha'}^{\;\;\alpha}F_{\alpha\beta} w^\beta = \frac{1}{2}
\epsilon_{\alpha'\beta} F^{\gamma\delta }F_{\gamma\delta} w^\beta=
\frac{1}{2} F^2 w_{\alpha'} =0 
\ee
Another important identity is given by
\be{rel2}
G^2 F_{\alpha\beta} = 0.
\ee
In order to prove this identity, we first multiply \eq{gf} by
$G_\delta^{\;\alpha\beta}$ to obtain
\be{intrel2}
\frac{1}{2} G^2 F_{\delta\sigma}  - 2
G_{\delta\beta\alpha}G^\alpha_{\;\sigma\gamma} F^{\gamma\beta} =0
\ee
The product $G_{\delta\beta\alpha} G^{\alpha}_{\;\sigma\gamma}$
contains both the spin 0 and  the spin 2 part. The spin 2 part
vanishes by \eq{gg}, therefore this 
product contains only the spin 0 part, 
which is given by
\be{spin0}
G_{\delta\beta\alpha} G^\alpha_{\;\sigma\gamma} =
-\frac{1}{6} \left( \epsilon_{\delta\sigma}\epsilon_{\beta\gamma} +
\epsilon_{\beta\sigma}\epsilon_{\delta\gamma}\right)  G^2 
\ee
Substituting the above equation in \eq{intrel2} gives \eq{rel2}.
We also have the identity 
\be{rel3}
F_{\alpha\beta} G^{\alpha\beta}_{\;\;\;\gamma} F^{\gamma\sigma} 
\equiv(F\cdot G)_\gamma F^{\gamma\sigma} =0 .
\ee
This equation is obtained by simply multiplying \eq{gf} by
$F_{\alpha\beta}$, the last two terms vanish due to the symmetry of
$G_{\alpha\beta\gamma}$ in all the indices. 
Finally,  using \eq{ring1} and \eq{ring2} we can show that
\be{rel4}
(F\cdot G)_\alpha w_\beta = -(F\cdot G)_\beta w_\alpha
\ee
The identities \eq{rel1}, \eq{rel2}, \eq{spin0} and \eq{rel3} 
imply that terms containing $G^2$ do not admit any 
expansions in higher powers of 
either $F$ or $G$. And the terms containing $(F\cdot G)$
also do not admit any higher powers of either $F$ or $G$. From these
considerations we can conclude that the only expansion which admits
arbitrary powers is an expansion purely in $F$.

\subsection{Chiral ring identities from closed string dual}

Though we do not have understanding of the chiral ring identities
discussed in the previous subsection we can appeal to the
open/closed string duality conjectured in \cite{Vafa:2000wi} 
to  derive them on the closed string side.
The basic conjecture of \cite{Vafa:2000wi}is that
${\cal N}=1$, $U(N)$ gauge theory with a single adjoint 
chiral superfield $\Phi$, realized, for instance, by $N$
D5-branes wrapped on a two-cycle of a local Calabi
Yau threefold,  is dual to closed string theory on
a CY threefold which is related by a conifold transition
to the previous one. On the closed string side 
${\cal N}=2$ supersymmetry
is broken down to ${\cal N}=1$ by the presence of
three-form fluxes on the CY space \cite{Taylor:1999ii}. The tree
level superpotential for $\Phi$ is related to the geometry
of the CY manifold. More precisely, the duality identifies
the lowest moment of the gauge theory operators of \eq{defoper}
with the components of an $U(1)$ ${\cal N}=2$ vector multiplet on the closed
string side  
\be{vec}
V(\hat{\theta}, \theta) = S(\theta) + \hat{\theta}^{\alpha} w_\alpha
(\theta)
+  \hat{\theta}^2 N.
\ee
$S$ is the closed string field dual to 
${\rm Tr}( W^\alpha W_\alpha) /32\pi^2$ , and it corresponds
to the complex structure modulus of the CY threefold, 
$w_\alpha$ is dual to 
${\rm Tr}( W_\alpha)/4\pi$  and $N$, 
the auxiliary field corresponding to the three-form flux
on the closed string side, is dual to 
${\rm Tr} (1)$ of the gauge theory. $\theta$ is the usual ${\cal N}=1$
superspace coordinate and $\hat{\theta}$ is the additional superspace
coordinate for ${\cal N}=2$ superspace. 
The duality is expected to still hold after coupling the gauge theory
on one side  and the vector multiplet on the other side,
to the supergravity background given by $F_{\alpha\beta}$
and $G_{\alpha\beta\gamma}$. In particular, a class of 
gravitational F-terms, usually called ${\cal F}_g$ 
\cite{Antoniadis:1994ze,Bershadsky:1993ta} 
are expected to match on the two sides,
with the above identification of fields. 

We can also organize the
background fields, the ${\cal N}=1$ Weyl multiplet and the graviphoton field
strength as an ${\cal N}=2$ Weyl multiplet as follows.
\be{2weyl}
H_{\alpha\beta} (\hat{\theta}, \theta) = F_{\alpha\beta} (\theta)+
\hat{\theta}^\gamma G_{\alpha\beta\gamma} (\theta)
\ee
In the above equation $F_{\alpha\beta}$ stands for the ${\cal N}=1$
self-dual graviphoton multiplet and 
$G_{\alpha\beta\gamma}$ refers to the ${\cal N}=1$
Weyl multiplet which contains the self-dual part of the Riemann
curvature. We have set the auxiliary field of the Weyl multiplet to
be zero as it does not play any role in deriving the ring relations
discussed in the previous sub-section. 

Our strategy to prove the basic ring relations in \eq{ring1}, 
\eq{ring2} and \eq{gf} would be to use the ${\cal N}=2$ Bianchi identities to
show that these equations are $\bar{D}$ exact. Here $\bar{D}$ refers
to the derivative of the ${\cal N}=1$ superspace  coordinate. This is the
same method used to obtain the ring relations in \eq{oring}.
Consider the following $\bar{D}$ exact quantity
\be{dex}
\bar{D}^{\dot{\alpha}} ( D_{\alpha\dot{\alpha}} V ) =  
[\bar{D}^{\dot{\alpha}}, D_{\alpha\dot{\alpha}} ] V
\ee
In writing the equality we have used the fact that $V$, the ${\cal N}=2$
vector multiplet, is annihilated by $\bar{D}$. From the definition of
covariant derivatives in superspace we have the following 
(see for instance \cite{Wess:1992cp} )
\bea{comcov}
( {\cal D}_C {\cal D}_ B &-& (-1)^{bc} {\cal D}_B {\cal D}_C ) 
V^{( A_1 A_2 \ldots )}  \\ \nonumber
&= &
- R_{CBD}^{A_1} V^{( D A_2 \ldots ) } 
- R_{CBD}^{A_2} V^{(A_1 D\ldots ) }  - \dots  
- T_{CB}^D {\cal D}_D V^{(A_1 A_2 \ldots)}
\eea
Here $A, B, C, D$, etc. refer either to 
bosonic or fermionic coordinates in
superspace, $b, c$ refer to their grading, a bosonic
coordinate having grade $0$ and a fermionic one grade $1$. 
$R$ and $T$ stand for the curvature and the torsion in superspace
coordinates respectively.  For ${\cal N}=2$ superspace,  the complete
solution for the Bianchi identities has been given in 
\cite{Howe:1982gz} and one can read out the required curvature and
torsion symbols. The equation in \eq{dex} can then be written as
\be{dexe}
\bar{D}^{\dot{\alpha}} (D_{\alpha\dot{\alpha} } V) = 
T^{\dot{\alpha} }_{\; \alpha\dot{\alpha}\gamma} \hat{D}^{\gamma} V
= H_{\alpha\gamma} \hat{D}^{\gamma} V
\ee
There are no curvature contributions as $V$ is a scalar in ${\cal N}=2$
superspace, there are other contributions to this Bianchi identity, but
they vanish on shell \footnote{This is the holomorphic counterpart of 
equation (7.6) in \cite{Howe:1982gz}. }. 
The zeroth and the first component in $\hat{\theta}$ reduces
to
\bea{r1r2}
F_{\alpha\gamma} w^{\gamma} &=& 0, \\
\nonumber
2 F_{\alpha\beta} N  + G_{\alpha\beta\gamma} w^{\gamma} &=& 0 
\eea
The $\hat\theta^2$ component is identically zero. 
These operator equations verify the lowest moment of the 
ring relations in \eq{ring1} and \eq{ring2} from the closed string
side. To verify all the moments of these relations we need map which
relates all the gauge invariant operators of the gauge theory to closed
string fields, at present such a detail map is lacking, though it is
obvious they will all be mapped to vector multiplets on the closed
string side.
To prove the relation in \eq{gf} consider the following $\bar{D}$
exact quantity
\bea{dex1}
\bar{D}^{\dot{\alpha}} ( D_{\alpha\dot{\alpha} } H^{\beta\gamma} ) &=&
[\bar{D}^{\dot{\alpha}}, D_{\alpha\dot{\alpha} } ] H^{\beta\gamma},
\\ \nonumber
&=& R^{\dot{\alpha} \beta}_{\;\;\;\alpha\dot{\alpha} \rho} H^{\rho\gamma} 
+  R^{\dot{\alpha} \gamma}_{\;\;\;\alpha\dot{\alpha} \rho} H^{\beta\rho} 
+ T^{\dot{\alpha} }_{\;\alpha\dot{\alpha}\rho} \hat{D}^{\rho}
H^{\beta\gamma}.
\eea
Substituting the required curvature and torsion symbols we get
\footnote{ We have used the holomorphic counterpart of equations (4.25)
and (7.6)  of \cite{Howe:1982gz}. }
\bea{dexp}
\bar{D}^{\dot{\alpha}} ( D_{\alpha\dot{\alpha} } H^{\beta\gamma} ) &=&
G^{\beta}_{\;\alpha\rho}H^{\rho\gamma} + 
G^{\gamma}_{\;\alpha\rho}H^{\rho\beta} + H_{\alpha\rho} 
\hat{D}^{\rho} H^{\beta\gamma}
\eea
There are other terms in this Bianchi identity, but they all vanish on
shell.  From the lowest component in $\hat{\theta}$ of the above equation we
obtain the relation \eq{gf}. Note that on lowering the $\beta$ and
$\gamma$ indices, the combination is entirely symmetric, thus containing
only the spin $3/2$ part of the tensor product of $G$ and $F$. 
This completes the proof of \eq{gf} from the closed string side.
The linear term in $\hat\theta$ of \eq{dexp} reduces to $\eq{gg}$

\section{Genus one analysis} 

The chiral
ring relations \eq{oring} ensures that only planar graphs contribute
to the computation of the superpotential in the absence of gravity or
the graviphoton field strength. 
The diagrammatic analysis of \cite{Ooguri:2003qp} and \cite{Ooguri:2003tt}
show that higher genus diagrams contribute when either gravity or the
graviphoton background is turned on. 
In fact the contribution of gravity alone enters at genus one in the
superpotential, and it was shown in 
\cite{David:2003ke} that the genus one correction to the loop equation
in the corresponding matrix model agrees with the gravitational
corrected anomaly equations in the gauge theory.
In this paper we extend this to the situation when the graviphoton
field strength is also turned on. The graviphoton affects the gauge
theory loop equations at all genera. In this section as an important 
preliminary step to the analysis at all genera
and, as a means to introduce all the definitions and methods, we will
analyze the anomaly equations of the gauge theory with the graviphoton
also turned on at genus one.  

To derive the Ward identities constraining the gauge invariant
generating functions 
of \eq{defoper} in the presence of gravity and the graviphoton field
strength we need three ingredients. 
Firstly,  the background modifies the ring to \eq{ring} and the
associated ring equations discussed in section 2.1 play a crucial
role.
The generalized Konishi anomaly \cite{Konishi:1984hf}
forms the second ingredient: one can
derive the Ward identities constraining the functions $R(z)$,
$w_\alpha(z)$ and $T(z)$ by considering an infinitesimal variation
$\delta \Phi_{ij}= f_{ij}$ where $f_{ij}$ is the matrix elements 
of the operators given in \eq{defoper}.
This variation is anomalous and, in absence of gravity, the anomaly is
given by
\begin{equation}
\frac{\delta f_{ji}}{\delta \Phi_{k \ell}} A_{ij,k \ell},
\end{equation}
with
\be{defa}
A_{ij,k \ell} = (W^2)_{kj} \delta_{i\ell} + \delta_{kj} (W^2)_{i\ell} - 2
W^{\alpha}_{kj} W_{\alpha i\ell}.
\label{an}
\ee
When a gravitational background is turned on, there is a direct
contribution to the Konishi anomaly This is just the generalized
gravitational contribution of the chiral anomaly 
\cite{Konishi:1988mb,Magnoli:1990qh}.
To include this
contribution we replace $A_{ij, kl}$ in \eq{defa} with 
\begin{equation}
A_{ij,k\ell} \rightarrow A_{ij, k\ell} + \frac{1}{3} G^2 \delta_{kj} \delta_
{i\ell}
\label{Gan}
\end{equation}
We expect that the presence of a graviphoton background will not
affect the Konishi anomaly. The graviphoton field strength is of  
dimension $3$, all terms in the Konishi anomaly equation are Lorentz
scalars and of dimension $3$. 
Thus there is no Lorentz invariant term which can be
constructed out of the graviphoton field strength which is of
dimension $3$. Therefore, the effect of the graviphoton  in the anomaly
equations can be seen only through the ring \eq{ring}.
Using these two ingredients, the equations determining the gauge
invariant operators of \eq{defoper} are given by
\bea{basan}
& &\langle R(z) R(z) \rangle  +\frac{1}{6} 
G^2 \langle w^\alpha(z) w_\alpha(z) \rangle 
-\langle {\rm Tr}( V'(\Phi){\cal R}(z)) \rangle   = 0, 
\nonumber\\
& &2\langle R(z) w_{\alpha}(z) \rangle -\frac{1}{3} G^2 \langle w_{\alpha}(z) 
T(z) \rangle   -\langle {\rm Tr}(V'(\Phi) \rho_{\alpha} (z)) \rangle   =0, 
\\
& &2\langle R(z) T(z) \rangle  -\langle {\rm Tr}(V'(\Phi){\cal T}(z)) \rangle  
\nonumber
+\langle w^{\alpha}(z) w_
{\alpha}(z) \rangle   
 - \frac{1}{3} G^2 \langle T(z) T(z) \rangle  =0.
\label{CanGF}
\eea
To arrive at these equations we have repeatedly used the identities in
the chiral ring derived in section 2.1. Had we not used those
identities, we would have found ambiguities 
in various terms, due to the fact that
cyclic property of the trace is not obeyed. 
Note that the above equations reduce 
to the same equations derived in \cite{David:2003ke} in absence of the
graviphoton fields strength. To see this, we have to  use the chiral ring
equation $G^2 w_\alpha =0$ in the first equation of \eq{basan}.
Finally, the third ingredient in solving for the gauge invariant
operators is that the above Ward identities involve two
point functions of the gauge invariant operators. In absence of either
gravity or the graviphoton field strength these operators factorize in
the chiral ring \cite{Cachazo:2002ry}. 
However, in the presence of these background
there is no apriori reason  for factorization, in fact the
correspondence of the gauge theory with the matrix model and the
diagrammatic calculations of \cite{Ooguri:2003qp,Ooguri:2003tt} imply
that these operators do not factorize. Therefore, we need a further set
of Ward identities determining the connected two-point functions. 
For the case of the gravitational background alone, this was done in
\cite{David:2003ke}, and it was shown there that the corrections to the gauge
invariant operator $T$ is precisely that of genus one correction to the
resolvent of the matrix model. We will
repeat the same analysis for the case where 
the graviphoton  field strength is turned
on. In section 3.1 we set up the anomaly equations which determine the
connected two point functions and in section 3.2 we will solve for the
one point functions of the gauge invariant operators, to genus one.

Before we proceed, we note that the
equations of \eq{basan} simplify  if we perform the
following field redefinition 
\begin{equation}
\label{redef1}
R(z) \rightarrow  {R}(z) +\frac{1}{6} G^2 T(z)
\end{equation}
This field redefinition shifts all moments in the generating
functional, and it is therefore a 
generalization of the field redefinition noted in 
\cite{David:2003ke}, which removed the genus zero contribution of the
gravitational correction to the superpotential. In fact, in section 3.2
we will see that this field redefinition does the same job in the
general case.
Using this redefinition in \eq{basan} we obtain
\bea{shfan}
\langle R(z) R(z) \rangle -I(z) \langle  R(z)) \rangle   = 0 
\nonumber\\
2\langle R(z) w_{\alpha}(z) \rangle  -I(z) \langle w_{\alpha}
(z) \rangle   =0 \\
2\langle R(z) T(z) \rangle  -I(z) \langle T(z) \rangle  
\nonumber
+\langle w^{\alpha}(z) w_
{\alpha}(z) \rangle  = 0
\eea
In obtaining these equations we have used the chiral ring identities
as well as \eq{basan}.
Here the integral operator $I(z)$  is given by
\be{intop}
I(z) A(z) = \frac{1}{2\pi i} \oint _{C_z} dy \frac{ V'(y) A(y)}{y-z} ,
\ee
with the contour $C_z$ encircling $z$ and $\infty$. Note that if $A$
is equal to ${\cal R} $, $\rho_\alpha$ or ${\cal T}$, the integral
operator reduces to
\be{defintop}
I(z) A(z) = V'(\Phi) A(z)
\ee
With the field redefinition above, 
the  Ward identities reduce exactly to those with no
gravitational or graviphoton bacground, in fact the first equation of
\eq{shfan} is identical to the equation for the matrix model
resolvent.

\subsection{Connected two point functions}

Following the same method used  in \cite{David:2003ke} 
we will derive a set of equations for the connected two point functions. 
We see that, to solve for the one point functions from \eq{shfan}, it is
sufficient to determine the connected two-point functions evaluated at 
coincident points in the $z$ plane. 
However, it is more convenient to determine 
the connected two point functions at two different points in the
complex plane say $z$ and $w$, for example $\langle R(z) T(w) \rangle$. 
This turns out be useful as we can impose conditions on the connected
two point function, such as their contour integrals around various
branch cuts in $z$ and $w$ plane vanish separately,
which enable us to solve the corresponding generalized Konishi
anomaly equations completely. 

We will briefly review the method used in \cite{David:2003ke} 
using the example of the two point function $\langle R(z) T(w)
\rangle$. Consider the infinitesimal transformation which is local in
superspace coordinates $(x^\mu, \theta, \bar\theta)$, 
\begin{equation}
\delta \Phi_{ij} = {\cal R}_{ij}(z)  T(w)
\end{equation}
The Jacobian of this transformation has two pieces
\be{ojac}
\frac{\delta (\delta \Phi_{ji})}{\delta \Phi_{kl}}= \frac{\delta
{\cal R}_{ji}(z)}
{\delta \Phi_{kl}} T(w) + \sum_m {\cal R}_{ji}(z) {\cal T}_{mk}(w) {\cal T}_{lm}(w)
\ee
The first term in the equation above together with the variation of
the classical superpotential  gives rise to 
\begin{eqnarray}
\langle (R(z) R(z) &~& - {\rm Tr}(V'(\Phi) {\cal R}(z) ) ) T(w)  \rangle  
\nonumber \\ \nonumber
&=&  \langle ( R(z) R(z) -{\rm Tr}(V'(\Phi) {\cal R} (z) ) )
\rangle \langle T(w) \rangle +
 2\langle R(z) \rangle 
\langle R(z) T(w) \rangle _c \\
&~& -\langle   {\rm Tr} (V'(\Phi) {\cal R}(z) ) ) T(w) \rangle _c 
+ \langle R(z) R(z) T(w) \rangle _c 
\label{c1}
\end{eqnarray}
where the subscript $c$ denotes completely connected 2- or 3-point
functions. 
The first term on the right hand side vanishes when we use the first
equation of \eq{shfan}.
From the second term in the Jacobian, when combined with the anomaly 
(\ref{an},\ref{Gan}), we obtain the following single trace contribution 
\begin{equation}
\label{c2}
-\frac{1}{3} G^2 \langle tr({\cal R}(z) {\cal T}(w) {\cal T}(w)) \rangle  
=- \frac{1}{3}
G^2 \partial_w \frac{\langle R(z)\rangle -\langle R(w)\rangle}{z-w}
\equiv -\frac{1}{3} G^2 \partial_w R(z,w)
\end{equation}
Here we have introduced the notation 
$A(z,w)=\langle A(z) - A(w)\rangle/(z-w)$ for 
$A$ equal to $R$, $w_{\alpha}$ and $T$. Note that the field
redefinition of \eq{redef1} does not affect the single trace
contribution, as it already comes with order $G^2$. The field
redefinition introduces a correction of order $G^4$ for the above
single trace quantity, which vanishes in the chiral ring.
Combining \eq{c1}, \eq{c2} and the first equation of (\ref{shfan}), 
one obtains the following equation for the 
connected correlation functions:
\be{rtc}
(2 \langle R(z)\rangle - I(z)) \langle R(z) T(w)\rangle_c  
+\langle R(z) R(z) T(w)\rangle_c 
-  \frac{1}{3}
G^2 \partial_w R(z,w) =0
\ee
Using estimates as the ones performed in \cite{David:2003ke},
shows that the completely connected three-point functions vanish.
We will not need these estimates when we discussing the
solution for all genera, as will be seen in the next section.
The method to obtain the other relevant connected correlation
functions is similar.  We need to 
consider a general variation of the form
\begin{equation}
\delta \Phi_{ij} = A_{ij}(z) B(w)
\end{equation}
where $A$ can be  ${\cal R}$, $\rho_{\alpha}$ or ${\cal T}$ and $B$ can be 
$R$, $w_\alpha$ or $T$. The resulting generalized Konishi anomaly
equations equations can be derived in the same way as above and 
are  written as the following  matrix equation.
\begin{eqnarray}
&~&
\left[   \begin{array}{ccc}
 M(z)  & 2\langle T(z) \rangle & 2 \langle w^\alpha(z) \rangle\\
0 & M(z) & 0 \\
0 & 2 \langle w_\alpha(z) \rangle & 
M(z)                      
\end{array} \right]
        \left[
\begin{array}{ccc}
\langle T(z)T(w)\rangle_c &\langle T(z)R(w)\rangle_c &\langle T(z)
w_{\beta}(w)\rangle_c \\
\langle R(z)T(w)\rangle_c &\langle R(z)R(w)\rangle_c &\langle R(z)
w_{\beta}(w)\rangle_c \\
\langle w_{\alpha}(z)T(w)\rangle_c &\langle w_{\alpha}(z)R(w)\rangle_c 
&\langle w_{\alpha}(z)
w_{\beta}(w)\rangle_c \end{array}\right]
= \nonumber\\ &~& =\partial_w
\left[
\begin{array}{ccc}
\frac{1}{3} G^2 T(z,w) & \frac{1}{3} G^2 R(z,w) & \frac{1}{3} G^2 w_ \beta(z,w)\\
\frac{1}{3} G^2 R(z,w) & 16 F^2 R(z,w) & -8 \left( F \cdot G \right)_\beta 
R(z,w) \\
 \frac{1}{3} G^2  w_\alpha (z,w)  &\,\,\, -8 \left( F \cdot G \right)_\alpha R(z,w) & Q_{\alpha \beta}
\end{array}\right]
\label{matrix}
\end{eqnarray}
Where we have introduced the operators 
$M(z) = (2\langle R(z)\rangle - I(z))$ and 
$Q_{\alpha \beta}=\frac{5}{3} G^2 \epsilon_{\alpha \beta} R(z,w) -8 
\left( F \cdot G \right)_\alpha  w_\beta (z,w)$, here the second term
is also proportional to $\epsilon_{\alpha\beta}$ using 
the chiral ring equation \eq{rel4}.
To obtain such equations we have used the chiral ring relations
extensively. We have also dropped all connected three-point functions.
The order at which they occur can be inferred using the 
estimates of \cite{David:2003ke}: they either vanish in the chiral ring
or occur at a higher order. 
For genus one we are interested in the
solution at order $G^2$, $F^2$ or $(F\cdot G)$.
The  equation in \eq{matrix} is of the form
\begin{equation}
{\cal M}(z) N(z,w) =\partial_w  K(z,w)
\end{equation}
with ${\cal M}(z)$  the first matrix operator appearing on the left hand
side of eq. (\ref{matrix}), $N(z,w)$ and $K(z,w)$ satisfy $N(z,w)=
N^t(w,z)$ and $K(z,w) = K^t(w,z)$. The non-trivial consistency
condition (integrability condition) is then given by
\begin{equation}
(\partial_w K(z,w)) {\cal M}^t(w) = {\cal M}(z) \partial_z K(z,w).
\label{int}
\end{equation}
Using the methods of \cite{David:2003ke}, it can be shown that the
above integrability condition is satisfied.
The existence of solutions for the connected two point functions, equation 
(\ref{matrix}), is guaranteed by the fulfillment of the integrability 
conditions (\ref{int}). However, as familiar also in 
matrix models, these solutions suffer from
ambiguities, in the form of a finite set of parameters. 
These ambiguities will be fixed by the physical requirement that the contour 
integrals around 
the branch cuts of the connected 
two point-functions, both in the $z$ and $w$ planes, vanish separately.
The reason for this is that the following operator equations hold:
\begin{equation}
\label{opequ}
\frac{1}{2\pi i}\int_{C_i}dz R(z) = S_i, ~~~~ \frac{1}{2\pi i}
\int_{C_i} dw T(w) = N_i, ~~~~\frac{1}{2\pi i}\int_{C_i}dz w_{\alpha}(z) =
w_{\alpha~i}.
\end{equation}
where $S_i$ is the chiral superfield whose lowest component is the
gaugino bilinear in the $i$-th gauge group factor in the broken phase
$U(N) \rightarrow  \prod_{i=1}^n U(N_i)$ and 
$w_{\alpha~i}$ is the $U(1)$ chiral gauge superfield of the $U(N_i)$
subgroup. Since these fields are background fields, in the connected
correlation functions the contour integrals around the branch cuts
must vanish. This requirement makes the solutions of the equations of
\eq{matrix} unique.

We now write the solution of all connected two-point functions in
terms of a single function.  Consider the equation
for the correlation function $\langle R(z) R(w) \rangle_c$ from the
matrix equation \eq{matrix}
\be{rt1}
M \langle R(z) R(w) \rangle_c - 16 F^2 \partial_w R(z,w) =0.
\ee
The above equation is a linear equation with an inhomogeneous term
which is proportional to $F^2$. Let the solution be given by
\begin{equation}
\label{rtsol}
\langle R(z) R(w) \rangle _c = -16F^2 H^{(1)}(z,w)
\end{equation}
where $H^{(1)}(z,w)$ solves the following equation \footnote{This
definition of $H$ differs from the one used in \cite{David:2003ke} by a
sign.}
\be{hsol}
M H^{(1)}(z,w) + \partial_w R(z,w) =0,
\ee
here the superscript in $H$ refers to the fact that we are working at
genus one.
It is possible to define such a function, as the inversion of operator 
$M$ is unambiguous. In 
\cite{David:2003ke} it was shown that 
the function $H(z,w)$ is symmetric in $z$ and $w$.  
We now illustrate how the  connected 
two point function $\langle T(z)T(w)\rangle_c$, 
can be expressed in terms of the function $H^{(1)}(z,w)$. 
From \eq{matrix}, the equation satisfied by this correlator is given by
\be{tt1eq}
M(z) \langle T(z)T(w)\rangle_c + 
2\langle T(z)\rangle  \langle R(z)T(w)\rangle_c
+ 2\langle w^\alpha(z)\rangle \langle w_\alpha(z) T(w) \rangle_c
= \frac{G^2}{3} \partial_w T(z,w)
\ee
Now let us define the following operators:
\begin{equation}
D = N_i \frac{\partial}{\partial S_i} , \,\,\,\,\,\,\, \delta_{\alpha}=w_{\alpha~i} \frac{\partial}{\partial S_i}
\end{equation}
Applying the operator $(D+ \delta^2/2)$ on equation \eq{hsol} we see
that it reduces to  \eq{tt1eq} if one uses the relations 
$\langle T(z)\rangle=(D+\frac{1}{2}\delta^2)\langle R(z)\rangle$ , and 
$\langle w_\alpha (z) \rangle = \delta_\alpha \langle R(z) \rangle$.
Though these relations are valid only to  the zeroth order,
it is possible to use them here 
since corrections occur at higher order than $G^2$ in \eq{tt1eq}.
Therefore, using the uniqueness of solutions of the equations
involving the operator $M$, we
find
\begin{equation}
\langle T(z)T(w)\rangle_c = - \frac{G^2}{3}(D +\frac{1}{2} \delta^2)H^{(1)}
\end{equation} 
We can find all other two point functions in a similar manner, at
genus one order they are given by
\begin{eqnarray}
\label{2pcon}
\langle R(z)R(w)\rangle_c = - 16 F^2 H^{(1)}, \,& &\,
\langle R(z)T(w)\rangle_c = - \frac{1}{3}G^2 H^{(1)}, \\ \nonumber
\langle w_{\alpha}(z) w_{\beta}(w)\rangle_c = 
(- \frac{5}{3} G^2 H^{(1)} + 8 F^2 D H^{(1)}) \epsilon_{\alpha \beta},
\,& &\,
\langle R(z) w_{\alpha}(w)\rangle_c = 8 (F \cdot G)_{\alpha} H^{(1)},
\\ \nonumber
\langle T(z) w_{\alpha}(w)\rangle_c = -\frac{1}{3} G^2 \delta_{\alpha} H^{(1)},
 & &
\langle T(z)T(w)\rangle_c = -\frac{1}{3}G^2(D +\frac{1}{2} \delta^2)H^{(1)}. 
\end{eqnarray}
 
\subsection{One point functions at genus one}

To solve for the corrections to the one point functions we first expand
the one point functions as
\begin{equation}
\langle R(z) \rangle = R^{(0)}(z) + R^{(1)}(z), \;\;
\langle w_\alpha(z) \rangle = w_\alpha^{(0)}(z) + w_\alpha^{(1)}(z), \;\;
\langle T(z) \rangle = T^{(0)}(z) + T^{(1)}(z),
\label{exp}
\end{equation}
where the terms with the superscript $0$  denote the zeroth order contribution 
in $F^2$, $G^2$ and $\left(F \cdot G\right)$ and terms with the superscript 
$1$ denote the first order contribution. 
Substituting the above expansions in \eq{shfan} we obtain
the following equations for the genus one contributions
\bea{goneq}
&~&(2R^{(0)}(z)-I(z))R^{(1)}(z)+16 F^2 H^{(1)}(z,z)=0 \nonumber\\ 
&~&(2R^{(0)}(z)-I(z))w_{\alpha}^{(1)}(z)-16 (F \cdot G)_{\alpha} H^{(1)}(z,z)+2R^{(1)}(z)w_\alpha^{(0)}(z) = 0
\\ 
&~&(2R^{(0)}(z)-I(z))T^{(1)}(z)+4 G^2 H^{(1)}(z,z)- \nonumber\\ 
&~& \hspace{1 in} -16 F^2 D H^{(1)}(z,z)+2 w^{(0) \, \alpha}(z) w_\alpha^{(1)}(z) +                2 R^{(1)}(z) T^{(0)}(z) 
=0 \nonumber
\eea
All the above equations are linear in the one point functions with
different inhomogeneous terms. The linear operator is $M = ( 2 R^{(0)} -
I(z))$, therefore we consider the equation
\be{oneptfo}
(2 R^{(0)}(z) - I(z)) \Omega^{(1)} (z)  = 
H^{(1)}(z,z)
\ee
From the definition of the operator $I(z)$ the solution of this
equation is given by
\be{finposo}
\Omega^{(1)}(z)=\frac{H^{(1)}(z,z)+c^{(1)}(z)}{2 R_0(z)-V'(z)},
\ee
here $c^{(1)}$ is the finite ambiguity in the solution, 
a polynomial of degree $n-2$. This ambiguity is again fixed by the
physical requirement that the contour integral of  $\Omega^{(1)}$, which
is proportional to the genus one correction to any one one-point of
interest, vanishes around branch cuts. This requirement ensures that
operator equations \eq{opequ} are valid and the background fields
$S_i$, $N_i$ and $w_{\alpha i}$ do not receive any $G^2$, $F^2$ or
$(F\cdot G)$ corrections.
Using these inputs, the genus one corrections to the one-point
functions of interest are given by
\bea{gonsol}
R^{(1)}(z) &=& 16 F^2 \Omega^{(1)}(z)\\ \nonumber
w_\alpha^{(1)}(z) &=& -16 (F \cdot G)_{\alpha} \Omega^{(1)}(z)\\
\nonumber
T^{(1)}(z) &=& 4 G^2 \Omega^{(1)}(z)-16 F^2 D \Omega^{(1)}(z)
\eea
At this juncture it is worthwhile to point out the difference in the
results had we not used
the field redefinition in \eq{redef1}. 
We would, in fact, be left with genus zero contributions, 
in addition to the corrections found in
\eq{gonsol}.
This is seen as follows:  without the field definition there will be
additional terms
proportional to $G^2$, multiplying products of one point functions
at the zeroth order. For example, in the last equation of \eq{basan}
there is a term 
proportional to $G^2 (T^{(0)}(z))^2$  
which, since it  goes like  $N^2$ , it represents a genuine 
genus 0 contribution,

\section{Solution at all genera}

In this section we obtain the complete solution for the one point
functions $R$, $w_\alpha$ and $T$ for all genera.
From \eq{shfan} we see that we need the connected two point functions
$\langle RR\rangle_c, \langle R w_\alpha\rangle_c, \langle R
T\rangle_c$ and 
$\langle w^\alpha w_\alpha\rangle_c$ to solve for the one point functions.
Our strategy in this section is to first obtain equations for
generating functionals for any arbitrary correlator using the generalized
Konishi anomaly.  This can be done by introducing sources coupled to
the operators of interest. 
The equations constraining the generating functionals
turn out to be a set of integro-differential
equations. From these we solve for the relevant two-point functions
which in turn enables us to 
obtain the one point-functions of interest. 
We then compare the results with the matrix model.
In the appendix we demonstrate that the integro-partial
differential equations constraining the generating functional 
are consistent and provide the details of the solution for all
connected two point functions.

\subsection{Generating functionals for connected correlators.}

To obtain equations for the generating functionals for connected
correlators we extend the method used to obtain equations for the
connected two point functions. Consider the following generating
functional for the operators of our interest
\be{genf}
\langle Z\rangle = \langle \exp 
\left[ \int dw \left( j_R(w)  R(w)  + j_{w}^\alpha (w) w_\alpha (w) + 
j_T(w) T(w) \right) \right] \rangle
\ee
The generating functional is a function of three variables $j_R,
j_w^\alpha, j_T$.There
are three equations which constrain $Z$ which are obtained by
considering the following three variations
\be{var}
\delta \Phi_{ji} = {\cal R}_{ji}(z) Z, \;\;\;
\delta \Phi_{ji} = \eta^\alpha \rho_{\alpha ji}(z) Z, \;\;\;
\delta \Phi_{ji} = {\cal T}_{ji} (z) Z,
\ee
here $\eta^\alpha$ is an arbitrary spinor.
We will now derive the constraint imposed by the Konishi anomaly for
the first variation. The derivation proceeds along the lines followed
to derive  the equations for the connected 
two-point functions, discussed in section 3.1.  
The Jacobian of the first variation in \eq{var}  is given by
\be{jac}
\frac{\delta (\delta \Phi_{ji} ) }{ \delta \Phi_{kl} } = 
\frac{\delta {\cal R}_{ji}  } {\delta \Phi_{kl} } Z + 
{\cal R}_{ji}(z) \frac{\delta Z}{\delta \Phi_{kl} }
\ee
where
\bea{jacz}
\frac{1}{Z} \frac{\delta Z}{\delta \Phi_{kl}} &=&
  \sum_m  \int dw   \left. 
j_R(w) {\cal R}_{mk}(w)  {\cal T}_{lm} (w) + 
j_w^\alpha (w) {w_\alpha}_{mk}(w)  {\cal T}_{lm} (w)  \right. \\ \nonumber
&+& \;\;\;\;\; \left.   j_T (w) {\cal T}_{mk}(w)  {\cal T}_{lm} (w)  \right)
\eea
Note that the expression contains terms similar to the Jacobians in 
\eq{ojac}.  Therefore, we can use the
anomaly equations constraining  the two-point functions
for obtaining the  equations constraining the
generating functional. 
The first term of the Jacobian in \eq{jacz} gives rise to
a term which is the product of
the first anomaly equation in \eq{shfan} times $Z$,
while  the second term of the Jacobian in \eq{jacz} gives rise to
inhomogeneous single trace terms similar to those in  \eq{c2}.
Using these considerations, the anomaly equation for the above
variation can be shown to reduce to
\bea{rgen}
& &\langle R (z) R(z) Z \rangle   
- I(z) \langle R(z) Z\rangle  \\ \nonumber
&-& \left\langle \int dw  \left(   16F^2 j_R \partial_w R(z, w)   
+ \frac{G^2}{3} j_T \partial_w R(z,w) - 8 j^\alpha_w (F\cdot G)_\alpha 
\partial_w R(z, w)
\right) Z
\right\rangle    = 0 
\eea
where $R(z, w) = ( R(z) - R(w) )/(z-w)$, $T(z, w) = (T(z) - T(w)) /(z-w)$.
We can write the above equation as an integro-differential equation by
introducing functional derivatives with respect to the sources 
on the generating functional. We
define
\bea{defder}
\partial_R^z =  \frac{\delta}{\delta j_R(z)}, \;\;\; \partial_T^z =
\frac{\delta}{\delta j_T(z)}, \;\;\; \partial_\alpha^z =
\frac{\delta}{\delta j_w^\alpha (z)}, \\ \nonumber
O_R^{(z,w)} = \partial_w \left( \frac{1}{z-w} \left[
\frac{\delta}{\delta j_R(z) } -\frac{\delta}{\delta j_R(w) } \right]
\right), \\ \nonumber
O_T^{(z,w)} = \partial_w \left( \frac{1}{z-w} \left[
\frac{\delta}{\delta j_T(z) } -\frac{\delta}{\delta j_T(w) } \right]
\right), \\ \nonumber
O^{(z,w)}_\alpha  = \partial_w \left( \frac{1}{z-w} \left[
\frac{\delta}{\delta j_w^\alpha(z) } -\frac{\delta}{\delta j_w^\alpha(w) } 
\right] \right). 
\eea
Now writing \eq{rgen} in terms of these derivatives gives
\be{rgenz}
\left[  \partial_R^2 - I(z) \partial_R  
- \int dw \left( 16F^2 j_R O_R  + \frac{G^2}{3} j_T O_R - 8
j_w^\alpha (F\cdot G)_\alpha  O_R \right)  \right] Z =0
\ee
here we have suppressed the superscripts $z, w$ in the derivatives for
clarity of notation.
The generalized Konishi anomaly constraints from 
the other two variations in \eq{var} are given by
\bea{rhotgen}
& &\left[ 
2\partial_R\partial_\alpha - I(z) \partial_\alpha 
\right. 
\\ \nonumber
& & \left.   + \int dw \left( 
  ( 8 j_R (F\cdot G)_\alpha 
+ \frac{5}{3} G^2 j_{w\alpha})  O_R 
-( \frac{G^2}{3} j_T 
+  8 j_w^\beta (F\cdot G)_\beta ) O_\alpha  \right)  \right] Z =0 
\\ \nonumber
& & \\ \nonumber
& &\left[ 
2\partial_R \partial_T  -I(z) \partial_T 
+ \partial^\alpha \partial_\alpha 
- \frac{G^2}{3} \int dw \left( j_RO_R + j_TO_T + j_w^\alpha O_\alpha
\right) \right] Z =0
\eea
The above equations form a set of closed integro-partial differential
equations which determine $Z$

For the connected two point functions of our interest it is sufficient
to consider the following ansatz for the generating functional
\be{anz}
Z = \exp(M) = \exp \left( M_R(j_R)  + j_T M_T(j_R)  + 
j^\alpha_w M_\alpha(j_R)  + \frac{1}{2}j^\alpha_w j^\beta_w
M_{\alpha\beta}(j_R)  \right).
\ee
Here  the product $j_T M_T$ is understood to mean 
$\int dz j_T(z) M_T(z)$, 
similarly for other products in the
above expression. 
In the appendix we will generalize this to include all two point
functions. Note that for the various cumulants in the ansatz we have
allowed an arbitrary dependence of $j_R$  while we allow only a finite
set of moments in $j_T$ and $j^\alpha_w$. As we will see subsequently,
it is possible that higher moments in $j_R$ are non vanishing, but
higher moments in the other currents truncate in the chiral ring. 
An indication of this is clear form the nature of the two point
functions at genus one in \eq{2pcon}. Only the $\langle R R\rangle_c$
correlator is a function of $F^2$ alone, all the others involve powers
of $G$, thus they will truncate at some order in the chiral ring.
With this ansatz the connected two-point functions of
interest are given  by
\bea{full2pt}
\langle R(z) R(w) \rangle_c = \partial_R^z \partial_R^w M_R (j_R), 
&\;& \langle R(z) w_\alpha(w) \rangle_c = 
\partial_R^z M_\alpha(w, j_R) , \\ \nonumber
\langle R(z) T(w) \rangle_c = \partial_R^z M_T (w, j_R), &\;&
\langle w_\alpha(z)  w_\beta (w) \rangle_c =  M_{\alpha\beta}(z,w, j_R).
\eea
The three equations of \eq{rgenz} and \eq{rhotgen} in terms of the
cumulant generating functional  $M$ become
\bea{cugen}
\\ \nonumber
& &(\partial_R M)^2 + 
\left[  \partial_R^2 - I(z) \partial_R   \right. \\ \nonumber
& & \left. - \int dw \left( 16F^2 j_R O_R  + \frac{G^2}{3} j_T O_R - 8
j_w^\alpha (F\cdot G)_\alpha  O_R \right)  \right] M =0
\\ 
\nonumber
\\
\label{cugen2}
& &  2\partial_R M\partial_\alpha M 
+ \left[ 
2\partial_R\partial_\alpha - I(z) \partial_\alpha 
\right. 
\\ \nonumber
& & \left.   + \int dw \left( 
  ( 8 j_R (F\cdot G)_\alpha 
+ \frac{5}{3} G^2 j_{w\alpha})  O_R 
-( \frac{G^2}{3} j_T 
+  8 j_w^\beta (F\cdot G)_\beta ) O_\alpha  \right)  \right] M =0 
\\ \nonumber
\\ 
\label{cugen3}
\\ 
\nonumber
&  & 2\partial_ R M \partial_T M + 
\partial^\alpha M \partial_\alpha M  \\ \nonumber
& & + \left[ 
2\partial_R \partial_T  -I(z) \partial_T 
+ \hat\partial^\alpha \hat\partial_\alpha 
- \frac{G^2}{3} \int dw \left( j_RO_R + j_TO_T + j_w^\alpha O_\alpha
\right) \right] M =0
\eea
As $M_R$ is only a function of $j_R$, 
the equation determining $M_R$ can be obtained from \eq{cugen}, by setting
$j_T=j_\alpha=0$. This is given by
\be{req}
(\partial_R M_R)^2 + 
\left[  \partial_R^2 - I(z) \partial_R  
- 16F^2 \int dw   j_R O_R \right] M_R =0
\ee
We will see in subsection 4.4 that the above equation is identical to
the generating functional equation for the resolvent of the matrix
model. In fact the $F^2$ expansion of the above equation can be
identified with the $1/\hat N^2$ expansion of the equation of the
resolvent of the matrix model.
We now find the solutions of the connected two-point functions of interest, the
analysis is similar to that of the genus one case in section 3. 
In fact, the
solution is a direct  generalization of that case.
To obtain the connected two point function $\langle RR\rangle_c$
differentiate \eq{cugen} by $\partial_R^w$, where $w$ refers to
another point in the complex plane and set $j_T= j_\alpha =0$. We
obtain the following equation
\bea{bas2}
( 2\partial_R^z  M_R -I(z) ) 
\partial_R^z \partial_R^w M_R 
+  (\partial_R^z)^2 \partial_R^w M_R 
-16 F^2 O_R^{(z,w)} M_R  \\ \nonumber
-16 F^2 \int dw' j_R O_R^{(z,w')} \partial_R^w M_R =0
\eea
This forms a basic equation out of which the solutions for the other
cumulant generating functions in \eq{anz} will be constructed. 
The equation is linear in $\partial_R^z\partial_R^w M_R$ with an inhomogeneous
proportional to $F^2$, therefore the solution is
proportional to $F^2$.  Let the solution of  of \eq{bas2} be given by
\be{rr}
\langle R(z) R(w) \rangle_c = \partial_R^z \partial_R^w M_R = -16F^2 H(z, w),
\ee
where $H(z,w)$ is the solution of the  following equation
\bea{defh}
( 2\partial_R^z  M_R -I(z) ) H(z, w) + \partial_R^z H(z, w)  \\
\nonumber
+ O_R^{(z, w)} M_R 
- 16 F^2 \int dw' j_R \partial_{w'} \left( 
\frac{ H(z, w) - H(w', w) }{z-w'}  \right) =0
\eea
By this definition $H(z,w)$ is symmetric in $z$ and $w$.
We have assumed that the solutions of these equations are unique. An
argument in favour of this is as follows. For the lowest order in the
genus expansion (setting $j_R=0$, and $\partial_R^z H(z,w)=0$) 
these equations reduce to the ones
which were studied in \cite{David:2003ke}. It was shown there that,  by 
demanding the vanishing of the integrals 
of the the various connected two-point
functions around the branch cuts both in the $z$ and $w$ plane, the
solution is unique. The equation in \eq{defh} is a generalization  of
those equations. One can envisage a generalization of those arguments
for these equations, proving that the solution of \eq{defh} is
unique.
To obtain the connected two point function
$\langle R(z) w_\alpha (w)\rangle_c$ 
we differentiate \eq{cugen} with respect to
$\hat\partial_\alpha^w $ and then set $j_T = j_\alpha =0$, to obtain
\bea{rhoeq}
( 2\partial_R^z  M_R -I(z) ) \partial_R^z M_\alpha(w) 
+ \partial_R ^z \partial_R^z M_\alpha(w)
+ 8 (F\cdot G)_\alpha O_R^{(z,w)} M_R  \\ \nonumber
-16 F^2 \int dw' j_R O_R^{(z,w')} M_\alpha(w)  =0
\eea
Again comparing \eq{defh} and \eq{rhoeq} we obtain
\be{rrho}
\langle R(z) w_\alpha(w)\rangle_c= 
\partial_R^z M_\alpha (w) =  8 (F\cdot G)_\alpha H(z, w) 
\ee
At this point one might wonder if the equation for the above
correlator obtained by differentiating \eq{cugen2} by $\partial_R^z$ 
will reduce to \eq{rhoeq}. In the appendix we will see this is indeed
the case and that 
the set of integro-differential equations in
\eq{cugen}, \eq{cugen2} and \eq{cugen3} is in fact consistent. 
In order to  obtain the correlator $\langle R(z) T(w)\rangle_c$ we 
differentiate 
\eq{cugen} by $\partial_T^w$ and then set $j_T= j_\alpha =0$, we
obtain
\bea{rteq}
( 2\partial_R^z  M_R -I(z) ) \partial_R^z M_T(w) 
+ \partial_R^z \partial_R^z M_T(w) 
- \frac{G^2}{3} O_R^{(z,w)} M_T(w) \\ \nonumber
-16 F^2 \int dw' j_R O_R^{(z,w')} M_T(w) =0
\eea
Comparing \eq{rteq} and \eq{defh} yields  
\be{rt}
\langle R(z) T(w) \rangle_c=  \partial_R^z M_T(w) = -\frac{G^2 }{3} H(z, w)
\ee
Now differentiate \eq{cugen2} with respect to $\partial_\beta^z$ and 
then set $j_\alpha, j_T=0$ to obtain the connected correlator $\langle
w_\alpha(z) w_\beta(w)\rangle_c$. We get
\bea{rhorhoeq}
& & ( 2\partial_R^z  M_R -I(z) )  M_{\alpha\beta}(z,w) 
+ 2 \partial_R^z M_\beta(w) M_\alpha(z) \\ \nonumber
&+& 2 \partial_R^z M_{\alpha \beta}  (z,w)  
- 8 ( F \cdot G)_\alpha \int dw' j_R O_R^{(z,w')} M_\beta(w) 
- \frac{5}{3} G^2 \epsilon_{\alpha\beta} O_R(w,z)  M_R \\ \nonumber
&-&8 (F\cdot G)_\beta \partial_w \left( \frac{M_\alpha(z)
-M_\alpha(w)}{z-w} \right) =0
\eea
The last term in the above equation can be written as
\bea{lrel}
-8 (F\cdot G)_\beta 
\partial_w \left( \frac{M_\alpha(z) -M_\alpha(w) }{z-w} \right)
& =& 8 \epsilon_{\alpha\beta} F^2
\partial_w \left( \frac{M_T(z) -M_T(w) }{z-w} \right), \\ \nonumber
&=& 8 \epsilon_{\alpha\beta} F^2 D O_R^{(z,w)}   M_R
\eea
In the last equality we have used the relation, 
$T = (D  + 1/2 \delta^2)R$ which is valid at the zeroth order 
and the chiral ring relation \eq{ring2} and \eq{rel4}.
The second term in \eq{rhorhoeq} contains $\partial_R^z M_\beta(w)$.
Substituting \eq{rrho} for this we and 
using the equation \eq{ring2} we see that this term vanishes. 
Using \eq{rrho} we see that term containing the integral in \eq{rhorhoeq} 
is proportional to $(F\cdot G)_\alpha (F\cdot G)_\beta$, which also vanishes
using \eq{gg}. For clarity we write down \eq{rhorhoeq} after dropping
these terms
\bea{rhorhod}
& &( 2\partial_R^z  M_R -I(z) )  M_{\alpha\beta}(z,w) 
+ 2 \partial_R^z M_{\alpha \beta}  (z,w)   \\ \nonumber
&- & \epsilon_{\alpha\beta} \frac{5}{3} G^2 O_R(w,z)  M_R
+ 8 \epsilon_{\alpha\beta} F^2 D O_R^{(z,w)} M_R =0 
\eea
There are two inhomogeneous terms in the above equation which motivates 
the following ansatz
\be{rhorho}
M_{\alpha\beta}(z, w) = \epsilon_{\alpha\beta} 
\left(  -\frac{5}{3} G^2 H(z, w)  + 8 F^2 D
H(z,w) \right) 
\ee
With this ansatz the term containing the connected three-point
function vanishes, because it contains the derivative $\partial_R^z$ which
contains an extra factor of $F^2$. This is seen as follows: \eq{defh}
is identical to the corresponding matrix model equation for the
connected two
point function of the resolvent, from a t'Hooft counting analysis,
the connected three point function is down by a factor of $1/\hat
N^2$, which in the gauge theory implies that there is an extra factor
of $F^2$ since  the $F^2$ expansion of \eq{req} and \eq{defh} is identical
to the $1/\hat N^2$  expansion of the matrix model.
Comparing \eq{defh}, we see that 
\eq{rhorho} solves the \eq{rhorhod}, as with this ansatz the connected
three point function in \eq{defh} also vanishes, since it occurs with an
extra factor of $F^2$.

\subsection{Solutions for the one point functions.}

Having obtained the required connected two point functions we 
can now solve for the corrections to the one point functions.
The analysis is identical to the genus one case. We first expand the
one point functions  about the zeroth order as 
\be{expan}
\langle R(z) \rangle = R^{(0)}(z)  + \tilde R(z) , \;\;\;
\langle w_\alpha(z) \rangle =  w_\alpha^{(0)}(z) + \tilde w_\alpha(z) , \;\;\;
\langle T(z) \rangle = T^{(0)}(z) + \tilde T(z).
\ee
here
$\tilde{R}$, $\tilde w $ and $\tilde T$ denote corrections to the
zeroth order solution.
Now substituting the above expansion in \eq{shfan} we obtain the same
equations as \eq{goneq} but with $H^{(1)}(z,z)$ replaced with 
the full connected two point function $H(z,z)$, which is the solution
of \eq{defh}. These are given below
\bea{coreq}
& &(2 R^{(0)} - I(z) ) \tilde R(z) -16F^2 H(z,z) = 0, \\ \nonumber
& &(2 R^{(0)} - I(z) ) \tilde w_\alpha(z) + 16 (F\cdot G)_\alpha H(z,z)
+ 2 \tilde{R} w_\alpha^{(0)}
= 0, \\ \nonumber
& &( 2 R^{(0)} - I(z) ) \tilde T(z) 
-4 G^2 H(z,z) + 16 F^2 D H(z,z) + 2 \tilde w^\alpha w_\alpha^{(0)}+ 
2T^{(0)} \tilde{R} = 0
\eea
The  corrections are given by
\bea{corr}
\tilde R(z) = 16 F^2 \Omega (z), \;\;\;\; \tilde w_\alpha = -16
(F\cdot G)_\alpha \Omega (z) , \\ \nonumber
\tilde T(z) = 4 G^2 \Omega (z) - 16 F^2 D\Omega (z)
\eea
where
\be{defom}
\Omega (z) =  \frac{1}{2 R^{(0)} - V'(z) } 
(H(z,z) + c(z) ) 
\ee
where $c(z)$ is a polynomial of degree $n-2$ and is uniquely
determined by the requirement that the contour integrals of
$\Omega(z)$ around each branch cut vanishes.
Note that the last term of the second equation in \eq{coreq} vanishes, 
because after substituting the solution for 
$\tilde{R}$, we see that that term is proportional to $F^2 w_\alpha$,
which  is trivial in the chiral ring.
In the section 4.4 we show that this is exactly the answer
obtained from the matrix model.

\subsection{Shift invariance of the anomaly equations}

In this section we show that we can assemble all the equations for the
generating functions given in \eq{cugen}, \eq{cugen2} and \eq{cugen3}
into one superfield equation by introducing the auxiliary fermionic
coordinate $\psi_\alpha$. 
We first assemble the loop variables as
\be{loop}
{\cal R} (z, \psi) = R(z) + \psi^\alpha w_\alpha(z) - \frac{1}{2} \psi^2
T(z)
\ee
Note that with this notation the generalized Konishi anomaly 
equations \eq{shfan}  is given by
\be{koncur}
\langle {\cal R}^2(z, \psi)\rangle  - I(z) \langle {\cal R}(z, \psi)\rangle 
=0
\ee
These generalized Konishi anomaly equations are the same as in the
case of ${\cal N}=1$ gauge theories in flat space with no graviphoton
background. In that case,
the connected two-point functions in \eq{koncur} vanish, 
and this shift symmetry of the equations implied that
the superpotenial could be written as the integral
\cite{Cachazo:2002ry}
\be{flatsup}
W_{\rm eff} = 
\int d^2\psi {\cal
F}_p \left( S_i + \psi^\alpha w_{\alpha i} -\frac{1}{2} \psi^2 N_i
\right) 
\ee
We would like to demonstrate that this is also true for the case of
${\cal N} =1$ theories studied here.

We first show that the equations for the connected correlators can be
written as a supermultiplet in the $\psi$ space.
The background fields, the ${\cal N} =1$ Weyl multiplet and the ${\cal N} =1$ 
spin $3/2$ multiplet can be assembled  as 
\be{shiftweyl}
{\cal H}_{\alpha\beta}  (\psi) =  F_{\alpha\beta} -\frac{1}{2} \psi^\gamma
G_{\alpha\beta\gamma}
\ee
To write the generating functional in the superspace $\psi$ we
introduce the multiplet of the sources as follows
\be{sourmul}
{\cal J}(z, \psi) = J_T + \psi^\alpha j_\alpha -\frac{\psi^2}{2} j_R.
\ee
Then the  generating functional can be written as
\be{fulgen}
Z = \exp \left( \int d^2 \psi dw  {\cal J} (w, \psi){\cal  R}(w, \psi) 
\right), 
\ee
here we have normalized $\int d^2 \psi \psi^2 = -2$.
To obtain the various correlators from the generating functional we
introduce the following derivative in the supermultiplet space
\be{dermul}
\partial_{\cal R}^z = 
\partial^z_R + \psi^\alpha \partial_\alpha^z - \frac{\psi^2}{2}
\partial_T^z
\ee
With this notation the set of equations \eq{cugen}, \eq{cugen2} and
\eq{cugen3} reduce to
\bea{supmet}
\left[ (\partial_{\cal R}^z )^2 - I(z)  \partial_{\cal R}^z
- \frac{2}{3}  
\hat{D}^2 ( {\cal H}^2 ) \int dw {\cal J}(w,\psi)  
O_{{\cal R}}^{(z,w)}
\right.  \\ \nonumber
\left. + 8 \int dw \hat{D}^2 ({\cal H}^{\alpha\beta} {\cal J}(w, \psi))  
{\cal H}_{\alpha\beta}
O_{\cal R}^{(z,w)} \right] Z =0
\eea
where 
\be{defhat}
\hat{D}^2 = \epsilon^{\alpha\beta}
\frac{\partial}{\partial \psi^\beta}  
\frac{\partial}{\partial \psi^\alpha}  
\ee
In \eq{supmet} $O_{\cal R}$ is defined using the super derivative of
\eq{dermul}. In deriving \eq{supmet} we have used the identities
\eq{ring1}, \eq{ring2} and \eq{rel4}.
This implies that the solutions of the loop equations can be written
in a shift invariant way. 
It is easy to verify this from the solutions we have found in the
previous subsection. The corrections to the one point functions can be
written as
\bea{corsup}
\tilde{\cal R}(\psi, z)   
&=& 16 {\cal H}^2 
\Omega( S_i + \psi^\alpha  w_{\alpha i} -\frac{1}{2} \psi^2 N_i, {\cal
H}^2 ), \\
\nonumber
&=& 16 F^2 \Omega(S_i, F^2 )  -16 \psi^\alpha (F\cdot G)_\alpha \Omega(S_i, 0 ) 
\\ \nonumber
&-&\frac{1}{2} \psi^2 \left( 4G^2 \Omega(S_i, 0) -16 F^2 D\Omega(S_i, 0) \right)
\eea
To obtain the expansion in the second line we have used the chiral
ring relation $F^2 w_{\alpha i}=0$ and $2 F_{\alpha\beta} N_i +
G_{\alpha\beta\gamma} w^\alpha_i =0$. Here the dependence of $F^2$ in
the last two terms are set to zero as the expansion in $F$ truncates in
the chiral ring for these terms.
Since the solutions to the correlators in ${\cal R}$ can be written as
a supermulitplet in $\psi$, the corrections to the super potential 
must be written as
\be{fullcsup}
\tilde W_{\rm eff} = \int d^2\psi {\cal H}^2 {\cal F} 
\left(S_i + \psi^\alpha w_{\alpha i}
-\frac{1}{2} \psi^2 N_i \right)
\ee
This motivates the completion of the gravitational corrections to the 
superpotential to as
\be{movcoms}
\tilde W_{\rm eff} = \int d^2\psi {\cal H}^{2g} {\cal F}_{g}
\left(S_i + \psi^\alpha w_{\alpha i}
-\frac{1}{2} \psi^2 N_i \right)
\ee
which is in agreement with what is obtained from the closed string
duality with $\psi$ playing the role of the second ${\cal N} =2$
superspace coordinate. However because of the identities in the chiral
ring all terms for $g>1$ are trivial form ${\cal N} =1$ point of view.
Nevertheless, the generating functional $\langle R(z)\rangle$ sees the
complete genus expansion and is identified with the matrix model  genus
expansion as we will see in the next subsection.

\subsection{Comparison with the matrix model results.} 

Consider a hermitian matrix model with an action given by 
\be{matact}
S = \frac{\hat N}{g_m} \sum_ k \frac{g_k}{k} {\rm Tr} M^k \equiv
\frac{\hat N}{g_m} V,
\ee
where $M$ is a hermitian $\hat{N} \times \hat {N}$ matrix.  The basic
loop equations for the resolvent is given by
\be{baslo}
\langle \Omega_m (z) \Omega_m(z)\rangle  -I (z) \langle \Omega_m(z) \rangle
=0
\ee
here $\Omega_m$ is the matrix model resolvent given by
\be{defomeg}
\Omega_m(z)  = \frac{g_m}{\hat N} {\rm Tr} \left( \frac{1}{z-M} \right)
\ee

We now obtain the loop equations satisfied by the variation 
\be{matv}
\delta M_{ji} = \Omega_{m\, ji} \exp\left( 
\int dw J(w)  \Omega_m(w)\right)  = \Omega_{m\, ji} Z_{m}
\ee
here $Z_{m}$ is the generating functional for the $n$-point
functions for the resolvent of the matrix model.
The loop equations for the variation in \eq{matv} is given by
\be{genlo}
\langle \left[ \Omega^2_m (z) -I(z) \Omega_m(z)   + (\frac{g_m}{\hat N})^2 
\int dw J \partial_w \left( \frac{ \Omega_m (z) - \Omega_m(w) }{z-w}
\right)  \right] Z_{m} \rangle =0
\ee
Writing this loop equation by introduction functional derivatives in
the current $J$ we get
\be{parlom}
\left[ \partial_J^2 - I(z) \partial_J
+ (\frac{g_m}{\hat N} )^2  \int dw J O_J  \right ] Z_m =0
\ee
To obtain equations for the connected correlators we introduce the
cumulant generating functional $Z_m = \exp ( M_m)$. The
cumulant generating functional satisfies the following equation
\be{cumat}
(\partial_J M_{m} ) ^2 + \left[
\partial_J^2 - I(z) \partial_J  + (\frac{g_m}{\hat N})^2 \int dw J O_J
\right] M_m =0
\ee
This equation is identical to the equation \eq{req} which is satisfied by 
cumulant generating functional $M_R$  of the gauge theory. The 
$F^2$ expansion in the gauge theory is analogous to the $1/\hat N^2$
expansion in the matrix model.  
The equation for the connected two point function is obtained by
differentiating the above equation with $\partial_J^w$, which is given
by
\bea{2ptm}
(2 \partial_J^z  M_m - I(z) ) \partial_J^z\partial_J^w M_m 
+ \partial_J^z\partial_J^z \partial_J^w M_m  
+ (\frac{g_m}{\hat N})^2 O_J^{(z,w)}  \\ \nonumber
+ (\frac{g_m}{\hat N} )^2 \int dw' J O_J^{(z, w')} \partial_J^w M_m
=0
\eea
By comparison with \eq{defh} we see that the solution of the
connected two point function of the matrix model resolvent is given by
\be{con2ptm}
\langle \Omega_m (z)  \Omega_m(w) \rangle = \partial_J^z \partial_J^w M_m
= (\frac{g_m}{\hat N})^2 H(z,w). 
\ee
Here again we have used the fact that the inversion of the operator
$(2 \partial_J^z M_m - I(z))$ is unambiguous if the ambiguity is fixed
by demanding contour integrals of the two point function around the
branch cuts vanish.
Note that in \eq{2ptm} using a t'Hooft counting analysis one finds that
the contribution of a $l$ loop planar Feynman graph to
the three point function $\partial_J^z\partial_J^z\partial_J^w M_m$ is
proportional to $(g_m/\hat N)^{(4+l)}$. Therefore its contribution is
subleading compared to the other two terms, this  
and the fact that \eq{cumat} and \eq{req} are identical was used in
the gauge theory analysis to drop certain three point functions in the
chiral ring.
To find the  the solution  of the one point function of the resolvent
we expand $\Omega_m (z) = \Omega^{(0}_m (z) + \tilde \Omega_m(z)$,
and substitute it in \eq{genlo}.
$\Omega^{(0)}_m$ is the contribution of the planar graphs to the
resolvent.  The correction $\tilde \Omega_m (z)$  satisfies the
following equation
\be{matcor}
( (2\Omega^{(0)}_m(z)   -I(z) ) \tilde \Omega_m(z) + 
( \frac{g_m} {\hat N} )^2 H(z,z) =0 
\ee
Demanding that the contour integral of $\tilde{\Omega}_m(z)$ 
around the branch cuts are vanishing ensures that
the correction is identical to \eq{corr}, therefore
proving the all-orders matching between gauge theory and matrix model
correlators claimed in the introduction.

\section{Conclusions}

In this paper we have analyzed the issue of gravitational
effective F-terms, resulting from the integration of a massive
adjoint scalar with arbitrary tree level superpotential
and coupled to ${\cal N}=1$ super Yang-Mills and supergravity,
with the non-trivial, non-standard $F_{\alpha\beta}$ background.
The main results we obtained are, firstly, the determination of
the chiral ring relations in the presence of both $F_{\alpha\beta}$
and $G_{\alpha\beta\gamma}$, thereby extending previous analysis
of \cite{David:2003ke}. Secondly, the proof that the gauge
theory one-point function $\langle R(z)\rangle$ coincides
to all orders with the matrix model one-point function
$\langle \Omega(z)\rangle$, provided one identifies the expansion
parameters on the two sides. As mentioned in the introduction,
this is equivalent to the statement that the gauge theory
correlators $\langle {\rm Tr}W^2 {\Phi}^k\rangle $
and the matrix model ones,  $\langle {\rm Tr}{M}^k\rangle $,
are identical to all orders in a genus expansion, after
the appropriate identification of the expansion parameters.  

Concerning terms in the gauge theory low energy
action, we have also argued that the gravitational 
effective superpotential can be written in the form
$\int d^2\psi {\cal H}^{2g}{\cal F}_g(S+\psi^\alpha w_\alpha
-\frac{1}{2}\psi^2 N)$, as expected from 
\cite{Vafa:2000wi,Ooguri:2003qp,Ooguri:2003tt},
with ${\cal F}_g$  related to a genus-$g$ matrix model
partition function. We have however observed that, due to the
chiral ring relations of section 2, the above superpotential
turns out to be trivial (i.e. $\bar D$-exact) from the ${\cal N}=1$
point of view for $g\geq 2$. Of course, the equality
between gauge theory and matrix model correlators noted above,
survives the chiral ring relations, since it involves only
powers of $F^2$ on the gauge theory side.

From the technical point of view, 
we have followed a strategy based on generalized anomaly
equations and chiral ring relations, as first exploited in
\cite{Cachazo:2002ry} for the non-gravitational case and then adapted
to the case of super Yang-Mills theory coupled to supergravity
for genus one in \cite{David:2003ke}.
As already stressed, one of 
the points of the present work was to find the correct
chiral ring relations in the ${\cal N}=1$ gauge theory
coupled to supergravity, with a background $F_{\alpha\beta}$:
whereas we cannot claim we have a
``microscopic'' gauge theoretic derivation of them, we find
the fact that our assumptions agree, for gauge invariant operators,
with the relations we obtain on the dual, closed string side,
rather remarkable.

As for some open problems, it should be interesting to extend
our approach to other matter representations and to the
other classical (or exceptional) groups, extending the analysis of
\cite{Alday:2003gb,Kraus:2003jv} which did not involve gravity.
Also, remaining in the context
discussed here, on the gauge theory side
we have used the resolvent-like operators, $R(z)$, $w_\alpha(z)$
and $T(z)$, but only their contour integrals around the branch
cuts in the $z$-plane
(which in the semiclassical limit become the critical points
of the superpotential),
$S_i$, $w_{\alpha i}$ and $N_i$
are identified with the closed string vector multiplets. It
would be interesting to see if there is a more general
correspondence, which, on the gauge theory side involves
higher powers of $\Phi$.

\acknowledgments
The work of J.R.D, E.G and K.S.N is supported in part by 
EEC contract EC HPRN-CT-2000-00148.

\appendix

\section{Other two point correlators and integrability conditions}

For completeness we derive the connected two point functions 
$\langle T(z) T(w)\rangle_c$ and $\langle w_\alpha(z) T(w) \rangle_c$.
These are not used in evaluating the full one-point functions, but
serve to demonstrate the consistency of the constraints on the
generating functional given by the equations \eq{cugen}, \eq{cugen2}
and \eq{cugen3}. 
For convenience we define the following moments 
\bea{defnm}
\partial_T^w \partial_\alpha^z M |_{j_T = j^\alpha_w =0} = 
M_{\alpha T} (z,w) =\langle
w_\alpha(z) T(w) \rangle_c, \\ \nonumber
\partial_T^w \partial_T^z M |_{j_T =j^\alpha_w =0}  = 
 M_{TT} (z,w) = \langle T(z) T(w) \rangle_c 
\eea
We proceed as in section 4.1, to determine the correlator $\langle
w_\alpha (z) T(w)$. We differentiate 
\eq{cugen2} by $\partial_T^w$ and then set $j_T = j_\alpha =0$ which
gives
\bea{rhoteq}
( 2\partial_R^z  M_R -I(z) ) M_{\alpha T}( z, w)
+ 2\partial_R^z M_T(w)  M_\alpha(z) 
+ 2\partial_R^z M_{\alpha T}( z, w) \\ \nonumber
+8 (F\cdot G)_\alpha  \int dw' j_R O_R^{(z, w')}  M_T(w) 
-\frac{G^2}{3} \partial_z \left( \frac{ M_\alpha(z) -M_\alpha(w)}{z-w}
\right) =0 
\eea
First note that the term containing the integral drops out as it
involves a $\partial_R^z M_T(w)$ which, using \eq{rt},
is proportional to $G^2 ( F\cdot G )$ and thus is  
trivial in the chiral ring. 
The solution of the above equation can also be related to the function 
$H(z,w)$, and  this can be seen as follows: differentiate \eq{defh} by
$\delta_\alpha$, to obtain
\bea{delh}
( 2\partial_R^z  M_R -I(z) ) \delta_\alpha H(z, w) + 
2\delta_\alpha\partial_R^z M_R H(z, w)  +
\partial_R^z \delta_\alpha H(z, w)  \\
\nonumber
- 16 F^2 \int dw' j_R \partial_z \left( 
\frac{ \delta_\alpha H(z, w) - \delta_\alpha H(w', w) }{z-w'}  \right) + 
\delta_\alpha O_R^{(z, w)} M_R =0
\eea
Now we note that for the one point functions we have the following relations
\be{deltrel}
\delta_\alpha \partial_R^z M_R  = M_\alpha(z), \;\;\;\; \delta_\alpha
O_R^{(z,w)} M_R = M_\alpha (z, w)
\ee
These relations are valid at the zeroth order, but it is sufficient
for our purpose as we will consider a solution which is proportional
to $G^2$ . Therefore the higher order corrections are trivial in 
the chiral ring. 
By comparing \eq{rhoteq} and \eq{delh} we see that 
the ansatz 
\be{rhot}
M_{T\alpha}(w,z) = - \frac{G^2}{3} \delta_\alpha H(z,w)
\ee
solves \eq{rhoteq}, as it reduces to \eq{delh}. 
To show this one uses \eq{deltrel}, the fact that the term containing
the integral vanishes in the chiral ring for both the equations and 
that the three point functions in \eq{rhoteq} 
and \eq{delh} vanish with
the ansatz of \eq{rhot}, as they come with a higher power of $F^2$.
Finally, to obtain the correlator $\langle T(z) T(w)\rangle_c$  
we take the derivative of \eq{cugen3} with respect to
$\partial_T^w$ and set $j_T = j_\alpha =0$, giving the following
equation
\bea{tteq}
\\
\nonumber
( 2\partial_R^z  M_R -I(z) ) M_{TT}(w, z)
+ 2 \partial_T^w\partial_R^z M_R M_T(z)  + 2 M^\alpha(z)
M_{\alpha T}(z,w)  
+ 2 \partial_R^zM_{TT}(z,w) 
\\ \nonumber
+ \partial_T^w \hat\partial^{z\alpha} \hat\partial^z_{\alpha}M |_{j_T,
j_\alpha =0} 
- \frac{G^2}{3} \int dw j_R O_R^{(w',z)} M_T(z)
-\frac{G^2}{3} \partial_w \left(\frac{M_T(z) - M_T(w)}{z-w} \right) 
 =0 
\eea
On substituting the value of $O_R^{(w',z)} M_T(z)$ from \eq{rt} in 
the term containing the integral, we see that it is  
proportional to $G^4$ and therefore trivial in the chiral ring.
We can also substitute the solutions of $\partial_T^w\partial_R^z M_R$ 
and $M_{TT}$ obtained in \eq{rt} and \eq{rhot} in the above equations.
With this substitution it is easy to see that the solution of \eq{tt}
equation can be written in terms of $H(z,w)$,
differentiating  \eq{defh} by the operator 
$D+ \delta^2/2$. We obtain
\bea{ddelh}
( 2\partial_R^z  M_R -I(z) ) ( D+ \frac{1}{2} \delta^2) H(z,w)
+ 2 (D+ \frac{1}{2} \delta^2 )\partial_R^z  M_R H(z,w)  \\ \nonumber
+ 2 \delta^\alpha \partial_R^z M_R
\delta_\alpha H(z,w)  + 
 (D + \frac{1}{2} \delta^2) \partial_R^z H(z,w) +
(D + \frac{1}{2} \delta^2 ) O_R^{(z,w)} M_R  \\ \nonumber
-16F^2 (D+ \frac{1}{2} \delta^2 ) \int dw' j_R \partial_{w'}
\left( \frac{ H(z,w) - H(w',w)}{z-w'} \right) =0
\eea
We also have the following relation at the zeroth order
\be{deltrel2}
( D + \frac{1}{2} \delta^2 ) \partial_R^z M_R = 
M_T(z) , \;\;\; 
\ee
Again it is sufficient to use the relation at the zeroth order, 
since we will be interested in a solution which is proportional to $G^2$,
therefore higher order corrections to the above relation vanish in
the chiral ring. From comparing \eq{tteq} and \eq{ddelh} 
and using the relations \eq{deltrel} and \eq{deltrel2}, we see that
the following ansatz satisfies \eq{tteq}
\be{tt}
M_{TT}(w,z) = -\frac{G^2}{3}( D + \frac{1}{2} \delta^2 ) H(z,w). 
\ee
Note that, with this ansatz, the terms containing the integral and the
three-point functions in both \eq{tteq} and \eq{ddelh} vanish, so that
they reduce to the same equation.

To verify that the equations constraining the generating functionals
are consistent we obtain the correlators $\langle R(z) w_\alpha(w)
\rangle_c$, $\langle R(z) T(w) \rangle_c$, $\langle w_\alpha (z)
T(w)\rangle_c$ 
by a different route and show that they lead to the
same results as discussed earlier. It is possible to obtain these
correlators by a different route,  as partial derivatives commute and
it is not obvious that the results for the correlators \eq{cugen},
\eq{cugen2} and \eq{cugen3} will be the same.

First consider the correlator $\langle R(z) w_\alpha (w) \rangle_c$,
in \eq{rhoeq} we had obtained it by differentiating \eq{cugen} by 
$\partial_\alpha^w$, we could also obtain the same correlator by
differentiating \eq{cugen2} by $\partial_R^w$. Here we verify that 
we get the same result. On performing the differentiation and setting
$j_T=j_\alpha =0$ we obtain
\bea{rhoreq}
&&( 2\partial_R M - I(z) ) \partial_R^w M_\alpha(z) 
+ 2 \partial_R^w\partial_R^z M_R M_\alpha(z)  \\ \nonumber
&&+ 2 \partial_R^w\partial_R^z M_\alpha(z)
+ \int dw' 8j_R (F\cdot G)_\alpha O_R^{(z,w')}  M_\alpha(w)
+ 8 ( F\cdot G)_\alpha O_R^{(z,w)} M_R =0
\eea
The following ansatz solves the equation \eq{rrho}
\be{rhor}
\partial_R^w M_\alpha (z) = 8 (F. G)_\alpha H(z,w).
\ee
To see this, note that with this
ansatz the term containing $\partial_R^w\partial_R^z M_R M_\alpha(z)$
is trivial in the chiral ring as it is proportional to $F^2 w_\alpha$,
the term containing the integral also vanishes. Therefore comparing
\eq{defh} and \eq{rhoreq} we see that the above ansatz satisfies the
latter equation.  The solution is consistent with the one obtained in
\eq{rrho} as $H(z,w)$ is a symmetric function in $z$ and $w$.
Consider the two point function $\langle R T \rangle_c$, we now obtain
this correlator 
by differentiating \eq{cugen3} with $\partial_R^w$ and verify that the
result is consistent with \eq{rt}. On performing the differentiation 
and setting $j_T=j_\alpha =0$ we obtain
\bea{treq}
&&( 2\partial_R - I(z) ) \partial_R^w M_T(z)
+ 2\partial_R^w\partial_R^z M_R M_T(z) 
+ 2 \partial_R^w M^\alpha(z)M_\alpha(z) \\ \nonumber
&&+ 2 \partial_R^w\partial_R^z M_T^z 
+ \partial_R^w M^\alpha_{\;\alpha}(z,z) 
- \frac{G^2}{3} \int dw' j_R O_R^{(z,w')} \partial_w M_R
-\frac{G^2}{3} O_R^{(z,w)} M_R =0
\eea
The combination of the second and third terms in the above equation is
trivial in the chiral ring as shown below
\bea{23zero}
 2\partial_R^w\partial_R^z M_R M_T(z) 
&+&  2 \partial_R^w M^\alpha(z)M_\alpha(z) \\ \nonumber
&=& 2 ( 16 F^2 M_T(z)  - 8 (F\cdot G) _\alpha M^\alpha(z) ) H(z,w) 
=0
\eea
Here we have used \eq{ring1}. Now 
comparing \eq{treq} and \eq{defh}, it is clear that the following
ansatz solves \eq{treq}
\be{tr}
\partial_R^w M_T(z) = -\frac{G^2}{3} H(z,w)
\ee
The terms containing the integral and the connected three point
functions are trivial in the chiral ring, with the ansatz in \eq{tr}
for both \eq{treq} and \eq{defh} as they come with a higher power of
$F^2$. This solution is consistent with
the one obtained in \eq{rt} because $H(z,w)$ is a symmetric function.
Finally  we consider the two-point function $\langle Tw_\alpha\rangle$,
we obtain this by differentiating \eq{cugen3} by
$\partial_\alpha^w$  and setting $j_T = j_\alpha =0$
\bea{troheq}
( 2\partial_R^z M_R - I(z) ) M_{\alpha T} (w,z)
+ 2\partial_R^zM_\alpha(w) M_T(z)
+ 2 M_{\alpha}^{\;\beta}(z,w)M_\beta(z)  \\ \nonumber
+ 2 \partial_R M_{T\alpha}(z,w) 
+ \partial_\alpha^w \partial^{\beta z}\partial_\beta M|_{j_T=j_\alpha
=0}
- \frac{G^2}{3} \int dw' j_R O_R^{(z,w')}M_\alpha(w)  \\ \nonumber
-\frac{G^2}{3} \partial_w \left(
\frac{M_\alpha(z) - M_\alpha(w)}{z-w} \right) =0
\eea
Again the second and third term of the above equation can be further
simplified in the chiral ring,  by substituting the correlators from
\eq{rrho} and \eq{rhorho}
\bea{23sim}
& & 2\partial_R^zM_\alpha(w) M_T(z)
+ 2 M_{\alpha}^{\;\beta}(z,w)M_\beta(z)  \\ \nonumber
&=& 2\left( 8(F\cdot G)_\alpha H(z,w) M_T(z) + \frac{5}{3} G^2 H(z,w)
M_\alpha(z) - 8 F^2 DH(z,w) M_\alpha(z) \right) \\ \nonumber
&=& -\frac{2}{3} G^2 M_\alpha(z) H(z,w)
\eea
Here we have used the chiral ring equations \eq{ring1} and \eq{ring2}
Now comparing \eq{troheq} and \eq{delh} we see that the the solution
can be written as
\be{trho}
M_{\alpha T}(z,w) = -\frac{G^2}{3} \delta_\alpha H(w,z)
\ee
The terms containing the three point functions and the integrals in
\eq{troheq} and \eq{delh} vanish with the above solution.
This concludes the proof of the integrability of the constraints on
the generating functional.

\bibliographystyle{utphys}
\bibliography{hgenus}

\end{document}